\documentstyle[12pt]{article}
\def\beq{\begin{equation}}
\def\eeq{\end{equation}}
\def\bea{\begin{eqnarray}}
\def\eea{\end{eqnarray}}
\def\bq{\begin{quote}}
\def\eq{\end{quote}}

\def\nnb{\nonumber}
\def\ga{\left(}
\def\dr{\right)}
\def\aga{\left\{}
\def\adr{\right\}}
\def\lb{\lbrack}
\def\rb{\rbrack}

\def\rar{\rightarrow}
\def\nnb{\nonumber}
\def\la{\langle}
\def\ra{\rangle}
\def\nin{\noindent}
\def\ba{\begin{array}}
\def\ea{\end{array}}
\def\bm{\overline{m}}
\begin{document}
\topmargin -1.5cm
\oddsidemargin -.5cm
\evensidemargin -1.0cm
\pagestyle{empty}
\begin{flushright}
PM 95/06
\end{flushright}
\vspace*{1cm}
\begin{center}
\section*{Model-independent determination of $\overline{m}_s$ \\
from $\tau$-like inclusive decays in $e^+e^-$ and\\
implications for the $\chi SB$-parameters }
\vspace*{0.5cm}
{\bf S. Narison} \\
\vspace{0.3cm}
Laboratoire de Physique Math\'ematique\\
Universit\'e de Montpellier II\\
Place Eug\`ene Bataillon\\
34095 - Montpellier Cedex 05, France\\
\vspace*{1.5cm}
{\bf Abstract} \\ \end{center}
\vspace*{2mm}
\noindent
We determine the strange quark running mass of the $\overline{MS}$-scheme
by $simulating$  $\tau$-$like$ inclusive processes for the
$old$ Das-Mathur-Okubo sum rule relating the $e^+e^-$ into $I=0$ and
$I=1$ hadrons total cross-sections data. We obtain to three-loop
accuracy: $\overline{m}_s$(1 GeV)=($196.7\pm 29.1$) MeV. By combining
this result with the pseudoscalar sum rule estimate
of $(\overline{m}_d +\overline{m}_u)$
and the standard current algebra values of the light
quark mass ratios, we deduce the $average$:
$\overline{m}_d$(1 GeV)=$(10.3\pm 1.5)$ MeV,
$\overline{m}_u$(1 GeV)=$(5.0\pm 1.5)$ MeV and
$\frac{1}{2}\la \bar uu +\bar dd \ra$ (1 GeV) $
\simeq -[(228.6\pm 9.1) \mbox{MeV}]^3$.
Using also our value of $m_s$, we update the estimates of the
$K'(1.46)$ and $K^*_0(1.43)$ decay
constants and of the (pseudo)scalar
two-point correlator subtraction constants.
 Then, we deduce a  deviation of about 34\% from kaon PCAC
and the ratio of the
{\it normal-ordered} condensates:
$\la \bar ss \ra /\la \bar uu \ra = 0.68^{+0.15}_{-0.29}$,
which confirm previous findings from QCD spectral sum rules.
Finally, using the recent value of the
$\overline{m}_b$ from the $\Upsilon$-sum rules,
we deduce the scale independent quark-mass ratio: $m_b/m_s=34\pm 4$.

\vspace*{3.0cm}
\begin{flushleft}
PM 95/06\\
April 1995
\end{flushleft}
\vfill\eject
\setcounter{page}{1}
 \pagestyle{plain}
\section{Introduction} \par
The determination of the strange quark running mass  is of prime
importance for low-energy phenomenology, for CP-violation
and for SUSY-GUT or other model-buildings.
A large number of efforts have been devoted to the determinations
of the light quark masses in the past since the time of current algebra
where
one has succeeded to fix the light quark mass ratios \cite{GASSER}.
Within
the advent of QCD, one has also been able to give a precise meaning for
the definition of the running quark masses within
the $\overline{MS}$-scheme \cite{FLO} and to predict their absolute
values using QCD spectral sum rules \cite{SNB} \`a la
SVZ \cite{SVZ}, in the pseudoscalar to two-\cite{PSEUDO,GASSER,PSEU}
and three-\cite{LARIN}-\cite{BIJ},\cite{SNB} loops,
the scalar \cite{SCAL,SCALB,JAMIN} and in
the vector \cite{REIND,SNB} channels, while more recently the strange
quark mass has been obtained from lattice simulations \cite{LATT}.
\nin
The advantage of the (pseudo)scalar channels with respect to the vector
one is clearly the fact that the
light quark masses are leading couplings in the sum rules analysis,
which make $a~priori$
their determinations quite reliable. However, one has to work at
some large values of the sum rules scale in order to escape advocated
instanton
contributions \cite{IOFFE}, which effects are not under good control and
even controversial. Unfortunately, at this large scale the
contributions of the higher state mesons to the sum rules analysis
become important and $model$-$dependent$, as we do not yet have
complete data in these
channels before the running of the $\tau$-charm factory machine, where
there is
a hope to measure the up and down quark masses from an interference
between
the pseudoscalar and axial-vector channels \cite{STERN}. The most recent
and updated sum rule determination of the strange quark mass comes
from the scalar channel \cite{JAMIN},
while the vector sum rule \cite{REIND,SNB} leads to a smaller
and inaccurate result.
\nin
In the following, we shall reconsider the vector channel by proposing a
new method for determining the strange quark mass. We shall also
use the result for an update of the estimates of different
chiral symmetry breaking parameters from the (pseudo)scalar
sum rules. In so doing, we
exploit the present $unexpected$ success of the $\tau$-decay inclusive
process for determining accurately the value of the QCD coupling
$\alpha_s$ \cite{BNP,LDP,LEDI}
and the stability of the results obtained for arbitrary low
mass-hypothetical heavy lepton by using the $e^+e^-$ into hadrons data
at low-energy \cite{PICH}. The power of $\tau$-decays
compared with existing QCD spectral sum rules resides mainly on the
existence of the threshold factor $(1-s/M^2_\tau)^2$, which gives a
double zero suppression
near the time-like axis where QCD is inaccurate and on the particular
$s$-structure of the expression of the decay rate which suppresses to
leading
order the contribution of the dimension $D=4$ condensates appearing in
the Operator Product Expansion (OPE).
\nin
Therefore, for the determination of the strange quark mass, we shall
$simulate$ $\tau$-$like$ inclusive decays for the processes
$e^+e^-$ into $I=0$ and $I=1$ hadrons, which have the advantage to be
model-independent as we have complete data of the spectral function
in the region of interest.
\section{The method and the QCD expressions}
We shall be concerned with the two-point correlator:
\beq
\Pi_{\mu\nu}(q^2)\equiv i\int d^4x~e^{iqx}\la 0|{\cal T}J_\mu(x)
J^\dagger_\nu(0)
|0\ra
=\ga q_\mu q_\nu-q^2g_{\mu\nu}\dr \Pi(q^2)
\eeq
built from the $SU(3)$-flavour components of the electromagnetic current:
\beq
J_\mu(x)=V^{(3)}_\mu (x)+\frac{1}{\sqrt{3}}V^{(8)}_\mu (x),
\eeq
where:
\beq
V^{(a)}_\mu \equiv \frac{1}{\sqrt{2}}\bar{\psi}\gamma_\mu \lambda_a
\psi(x),~~~ a=3,8,
\eeq
and $\lambda_a$ are the diagonal flavour SU(3) matrices:
\beq
\lambda_3=\frac{1}{\sqrt{2}}
\ga
\ba{ccc}
1 & & \\
  &-1& \\
  & &0 \\
\ea
\dr,
{}~~~
\lambda_8=\frac{1}{\sqrt{6}}
\ga
\ba{ccc}
1 & & \\
  &1& \\
  & &2 \\
\ea
\dr,
\eeq
acting on the basis of the up, down and strange quarks:
\beq
\psi \equiv
\ga
\ba{c}
\psi_u\\
\psi_d\\
\psi_s\\
\ea
\dr.
\eeq
The trace over the colour degree of freedom of the quark fields is
understood. At the one-photon approximation, the $e^+e^-$ into hadrons
total cross-section is related to the absorptive part of the correlator,
via the optical theorem, as:
\beq
\sigma(e^+e^- \rar hadrons)=\frac{4\pi^2\alpha}{q^2}e^2\frac{1}{\pi}
\mbox{Im}\Pi(q^2),
\eeq
where:
\beq
-(g_{\mu\nu}q^2-q_\mu q_\nu)\mbox{Im}\Pi(q^2)
=\frac{1}{2}\int d^4x~e^{iqx}\la 0|{\cal T}\lb J_\mu(x),J^\dagger_\nu(0)
\rb |0\ra,
\eeq
and:
\beq
\alpha^{-1}=137.04
\eeq
is the fine electromagnetic structure constant. It is also convenient
to introduce the ratio of cross-section:
\beq
R^{(I)}\equiv\frac{\sigma(e^+e^-\rar I
\mbox{ hadrons})}{\sigma(e^+e^-\rar \mu^+\mu^-)}
\eeq
where:
\beq
\sigma(e^+e^-\rar \mu^+\mu^-)=  \frac{4\pi^2\alpha^2}{3q^2}.
\eeq
Using CVC, we
also know  that the vector component of the $\tau \rar {\nu}_\tau$ +
hadrons semi-inclusive width
can be related to the isovector $I=1$ component of the $e^+e^-$
cross-section as \cite{GIL}:
\beq
R_{\tau,I}\equiv \frac{3\cos^2{\theta_c}}{2\pi\alpha^2}S_{EW}
\int_{0}^{M^2_\tau}ds~\ga 1-\frac{s}{M^2_\tau}\dr^2\ga 1+\frac{2s}
{M^2_\tau}\dr
\frac{s}{M^2_\tau}~\sigma^{(I)}_{e^+e^-\rar~I},
\eeq
where $S_{EW}=1.0194$ is the electroweak correction
from the summation of the leading-log contributions \cite{MARC}.
In the following, we shall generalize this expression and
propose a $\tau$-$like~sum ~rule$ for the $I=0$ channel.
\nin
The QCD expression of the $I=1$ rate has been already derived \cite{BNP}
and reads:
\beq
R_{\tau,I}=\frac{3}{2}\cos^2{\theta_c}S_{eW}\ga 1+\delta_{EW}+\delta^{(0)}
+\sum_{D=2,4,...}{\delta^{(D)}_{ud,I}}\dr.
\eeq
$\delta_{EW}=0.0010$ is the electroweak correction coming from the
constant term \cite{LI}.
\nin
The perturbative corrections read \cite{BNP}:
\beq
\delta^{(0)}_I= \ga a_s\equiv \frac{\alpha_s(M_\tau)}{\pi}\dr+5.2023
a_s^2+26.366a_s^3+...,
\eeq
\nin
where the expression of the running coupling to three-loop accuracy is:
\bea
a_s(\nu)&=&a_s^{(0)}\Bigg\{ 1-a_s^{(0)}\frac{\beta_2}{\beta_1}\log
\log{\frac{\nu^2}{\Lambda^2}}\nnb \\
&+&\ga a_s^{(0)}\dr^2\lb\frac{\beta_2^2}{\beta_1^2}\log^2\
\log{\frac{\nu^2}{\Lambda^2}}-\frac{\beta_2^2}{\beta_1^2}\log
\log{\frac{\nu^2}{\Lambda^2}}-\frac{\beta_2^2}{\beta_1^2}
+\frac{\beta_3}{\beta_1}\rb+{\cal{O}}(a_s^3)\Bigg\},
\eea
with:
\beq
a_s^{(0)}\equiv \frac{1}{-\beta_1\log\ga\nu/\Lambda\dr}
\eeq
and
$\beta_i$ are the  ${\cal{O}}(a_s^i)$ coefficients of the
$\beta$-function in the $\overline{MS}$-scheme for $n_f$ flavours:
\bea
\beta_1&=&-\frac{11}{2}+\frac{1}{3}n_f \nnb \\
\beta_2&=&-\frac{51}{4}+\frac{19}{12}n_f\nnb \\
\beta_3&=&\frac{1}{64}\Big{[}-2857+\frac{5033}{9}n_f-\frac{325}{27}
n^2_f\Big{]}.
\eea
For three flavours, we have:
\beq
\beta_1=-9/2,~~~~\beta_2=-8,~~~~\beta_3=-20.1198.
\eeq
In the case of flavour-neutral current of interest here,
the $m^2_s$-term of the $D$-function is given by \cite{BNP,CHET}:
\beq
\Delta\Pi(Q^2)\simeq -\frac{3}{4\pi^2}
\frac{\overline{m}_s^2}{Q^2}\aga 1+\frac{8}{3}a_s
+25.6a_s^2 \adr,
\eeq
which leads to the mass corrections:
\beq
\delta_{ii,V}^{(2)}=-12\ga 1+\frac{11}{3}a_s
+K_3a^2_s\dr\frac{\bm^2_i}
{M^2_\tau},
\eeq
where $K_3\approx 20$ contains not only the $a_s^2$-coefficient of the
correlator but also terms induced from the Cauchy integration of lower
order terms;
$\bm_i$ is the running mass of the quark of flavour $i$ evaluated at
$M_\tau$.
The expression of the running quark mass in terms of the
invariant mass $\hat{m}_i$ is \cite{FLO}:
\bea
&&\bm_i(\nu)=\hat{m}_i\ga -\beta_1 a_s(\nu)\dr^{-\gamma_1/\beta_1}
\Bigg\{1+\frac{\beta_2}{\beta_1}\ga \frac{\gamma_1}{\beta_1}-
 \frac{\gamma_2}{\beta_2}\dr a_s(\nu)\nnb \\
&&+\frac{1}{2}\Bigg{[}\frac{\beta_2^2}{\beta_1^2}\ga \frac{\gamma_1}
{\beta_1}-
 \frac{\gamma_2}{\beta_2}\dr^2-
\frac{\beta_2^2}{\beta_1^2}\ga \frac{\gamma_1}{\beta_1}-
 \frac{\gamma_2}{\beta_2}\dr+
\frac{\beta_3}{\beta_1}\ga \frac{\gamma_1}{\beta_1}-
 \frac{\gamma_3}{\beta_3}\dr\Bigg{]} a^2_s(\nu)+
{\cal{O}}(a_s^3)\Bigg\},
\eea
where $\gamma_i$ are the ${\cal{O}}(a_s^i)$ coefficients of the
quark-mass anomalous dimension for $n_f$ flavours:
\bea
\gamma_1&=&2, \nnb \\
\gamma_2&=&\frac{101}{12}-\frac{5}{18}n_f,\nnb \\
\gamma_3&=&\frac{1}{96}\Big{[}3747-\ga 160\zeta\ga 3\dr +\frac{2216}{9}
\dr n_f
-\frac{140}{27}n^2_f\Big{]},
\eea
and $\zeta (3)=1.2020569...$is the Riemann zeta function.
\nin
For three flavours, we have:
\beq
\gamma_1=2,~~~~\gamma_2=91/12,~~~~\gamma_3=24.8404.
\eeq
The $D=4$ contributions read \cite{BNP}:
\beq
\delta^{(4)}_{ii,I}=\frac{11}{4}\pi a_s^2\frac{\la\alpha_s G^2\ra}
{M^4_\tau}
-36\pi^2a_s^2\frac{\la m_i\bar{\psi}_i\psi_i\ra}{M^4_\tau}-
 8\pi^2a_s^2\sum_k{\frac{\la m_k\bar{\psi}_k\psi_k\ra}{M^4_\tau}}
+36 \frac{\bm_i^4}{M^4_\tau},
\eeq
which due to the Cauchy integral and to the particular $s$-structure of
the inclusive rate, the gluon condensate and the linear terms in $m_i$
starts at ${\cal O}(a_s^2)$. This is a great advantage compared with the
ordinary sum rule for the vector current discussed in \cite{REIND,SNB}
as in that case there is a strong correlation between these two terms.
We shall use \cite{BNP,SNB}\footnote{A forthcoming analysis \cite{SNE}
gives a more precise value of the gluon and four-quark condensates:
$\la\alpha_s G^2\ra=(7.1\pm 0.9)10^{-2}~\mbox{GeV}^4,~
\rho\alpha_s\la \bar{\psi_i}\psi_i\ra^2=(5.8\pm 0.9)10^{-4}~\mbox{GeV}^6$.
 However, for the quantities which we shall use, their contributions
will vanish to leading order in the chiral symmetry breaking expansion.}
$^,$
\footnote{Here and in the following:
$ \la \bar ii \ra \equiv\la\bar{\psi}_i\psi_i\ra $.}
:
\bea
\la\alpha_s G^2\ra&=&(0.06\pm 0.03)~\mbox{GeV}^4\nnb \\
(m_u+m_d)\la \bar uu
+\bar dd\ra &=&-{2}m_\pi^2 f_\pi^2 \nnb \\
 (m_s+m_u)\la\bar ss+ \bar uu\ra&\simeq &
-{2}(0.5\sim 1)m_K^2 f_K^2,
\eea
where: $f_\pi=93.3$ MeV, $f_K=1.2 f_\pi$ and we have taken into account
a possible violation of kaon PCAC of about 50 $\%$ as suggested by the
QSSR analysis \cite{SNP,SNB} and which we shall test later on.
However, due to the $a_s^2$ suppression of
the $D=4$ contributions, our results will not be affected in a sensible
way by the exact value of these condensates.
\nin
The $D=6$ contributions read \cite{BNP}:
\beq
\delta_{ii,I}^{(6)}
\simeq 7\frac{256\pi^3}{27}\frac{\rho\alpha_s\la \bar{\psi_i}\psi_i
\ra^2}{M^6_\tau},
\eeq
where $\rho$ measures the deviation from the vacuum
saturation estimate of the four-quark condensate. We shall
use \cite{BNP,SNB}:
\beq
\rho\alpha_s\la \bar{\psi_i}\psi_i\ra^2=(3.8\pm 2.0)10^{-4}~\mbox{GeV}^6.
\eeq
In our next analysis, we allow that  $\la \bar ss \ra^2 \simeq
(\frac{1}{4}\sim 1)\la \bar uu \ra^2 $, consistently with the
previous possible violation of kaon PCAC.
\nin
The contribution of the $D=8$ operators in the chiral limit reads
\cite{BNP}:
\beq
\delta^{(8)}_{ii,I}\approx -\frac{39\pi^2}{162}\frac{\la \alpha_s G^2
\ra^2}{M^8_\tau},
\eeq
which can be safely neglected due to the high-power suppression in the
$\tau$-mass.
\begin{table*}[h]
\setlength{\tabcolsep}{1.5pc}
\newlength{\digitwidth} \settowidth{\digitwidth}{\rm 0}
\catcode`?=\active \def?{\kern\digitwidth}
\caption{Different contributions to $R_{\tau,0}$ }
\begin{tabular}{c|c c c  c}
\hline
&&&& \\
$M_{\tau}$[GeV]&$\omega(782)$
& $\phi(1019.4)$ & $\omega'(1419)$ & $continuum$ \\
&&&& \\
\hline
&&&& \\
$1.0$ & $0.160 \pm 0.01$ & $$ & $$ &
$0.003\pm 0.001$ \\
$1.2$ & $0.205 \pm 0.01$ & $0.19 \pm 0.01$ &  &$0.011\pm 0.003$
\\
$1.4$ & $0.189 \pm 0.01$ & $0.33 \pm 0.02$ & &$0.008\pm 0.002$
\\
$1.6$ & $0.160 \pm 0.01$ & $0.36 \pm 0.02$ & $0.01 $ &
$ 0.012 \pm 0.006 $\\
&&&& \\
\hline
\end{tabular}
\end{table*}
\section{Model-independent determination of $\overline{m}_s$}
\nin
In so doing, we can work with the $\tau$-like rate into $I=0$:
\beq
R_{\tau,0}
\eeq
 or with the
ratio of the $I=0$ over the $I=1$ rates
or with their difference:
\beq
\Delta_{10}\equiv R_{\tau,1}-3{R_{\tau,0}}~.
\eeq
This sum rule vanishes in the flavour $SU(3)$-limit $m_i=m_j$ and $
\la\bar{\psi}_i\psi_i\ra=\la\bar{\psi}_j\psi_j\ra$. In this respect, it
is historically very similar to the old Das-Mathur-Okubo sum
rule \cite{DMO} and its QCD version \cite{QDMO}.
The main advantage of these new $\tau$-like sum rules, compared with the
previous ones, is the presence of the $\tau$-threshold
factor which suppresses the contribution near the time-like axis where
the QCD expression is inaccurate. Here, one has
also a suppression of the leading order
$D=4$ condensate contributions, such that one can have a clean effect
from the quark mass corrections.
\nin
The QCD expression of $R_{\tau,0}$ and of $\Delta_{10}$
can be deduced from the previous formulae.
However, in order to have a faster convergence of the
perturbative series at the scale we shall work, we shall instead expand
the width $R_{\tau,I}$ in terms of the contour coupling \cite{LDP}
\footnote{We shall see in \cite{SNE} that, in the case of
$\tau$-decays, this procedure gives a good
approximation of the optimized perturbative series if higher order terms
are resummed.}:

\beq
A^{(n)}=\frac{1}{2i\pi}\oint_{|s|=M^2_\tau} \frac{ds}{s}\ga 1-2\frac{s}
{M^2_\tau}+2\frac{s^3}{M^6_\tau}-\frac{s^4}{M^8_\tau}\dr~a^n_s.
\eeq
The DMO-like sum rule reads:
\beq
\Delta_{10}=\frac{3}{2}\cos^2{\theta_c}~S_{EW}
\frac{2}{3}\sum_{D=2,4,...}{\ga \delta^{(D)}_{uu}-\delta^{(D)}_{ss}\dr}~,
\eeq
The phenomenological parametrization of $R_{\tau,0}$ and $\Delta_{10}$
is done using the available data of the $e^+e^-\rar$ hadrons total
cross-section below 1.6 GeV, where the $I=0$ part is measured with a
quite good accuracy thanks to the clean separation of the $I=1$
(even number of pions) over the $I=0$ (odd numbers) and of the
$\bar{K}K$ states originating from the $\rho$ or the $\phi$ resonances.
\begin{table*}[h]
\setlength{\tabcolsep}{1.5pc}
\caption{Phenomenological estimates of $R_{\tau,I}$ and $\Delta_{10}$ }
\begin{tabular}[h]{c|c c c  }
\hline
&&& \\
$M_{\tau}$[GeV]&$R_{\tau,0}$
& $R_{\tau,1}$ & $\Delta_{10}$ \\
&&& \\
\hline
&&& \\
$1.0$ & $0.160 \pm 0.010$ & $1.608\pm 0.064$ &
$1.128\pm 0.071$ \\
$1.2$ & $0.406 \pm 0.014$ & $1.900 \pm 0.075$ &  $0.682\pm 0.086$
\\
$1.4$ & $0.527 \pm 0.022$ & $1.853 \pm 0.072$ & $0.272\pm 0.098$
\\
$1.6$ & $0.542 \pm 0.023$ & $1.793 \pm 0.070$ &
$ 0.167 \pm 0.098 $\\
&&& \\
\hline
\end{tabular}
\end{table*}

\nin
For the $I=0$, we take
the parameters of the narrow $\omega$ and $\phi$-resonances from PDG94
\cite{PDG} to which we add the contribution of the $\omega'(1419)$. The
narrow width approximation gives an accurate estimate of these resonance
effects:
\beq
\sigma(e^+e^-\rar V)=12\pi^2 \frac{\Gamma_{V\rar e^+e^-}}{M_V}
\delta(s-M^2_V).
\eeq
Threshold effects due to $3\pi$ can be incorporated by using a
Breit-Wigner form of these narrow resonances or chiral perturbation
theory similarly to the one done for the $3\pi$ in the case of the
pseudoscalar channel \cite{BIJ}, but these effects are negligible
here. We have also estimated the
 continuum contribution outside the resonances by using a
least-square fit of the data compiled in \cite{DOL}, where we take the
largest range of experimental errors in order to have a conservative
estimate.
For 0.83 GeV $\leq \sqrt{s}\leq$ 0.99 GeV, the continuum effect to
$R_{\tau,0}$ is much smaller than the errors induced by the resonance
contributions. The same feature also happens for the continuum in the
range 1.11 GeV $\leq \sqrt{s}\leq$ 1.40 GeV. We give in Table 1 our
estimate of the different contributions to $R_{\tau,0}$.
\nin
$R_{\tau,1}$ will be estimated in the same way as in \cite{PICH} and will
be the same data points as the ones in the Fig. 1 of that paper. We
give the results for $R_{\tau,1(0)}$ and $\Delta_{10}$ in Table 2.
Using the phenomenological values of
$R_{\tau,0}$ and $\Delta_{10}$ given in Table 2 and their
QCD expressions presented previously, we extract the value of the running
masses for each value of $M_\tau$. The analysis from $R_{\tau,0}$ is not
conclusive. This is mainly due to the competition between the
perturbative radiative and quark mass corrections. The analysis from
$\Delta_{10}$
is given in Table 3,
where we show the value of
the invariant mass $\hat{m}_s$ obtained at different energy scale
for a given value of $\Lambda_3$ for three flavours.  We consider as a
final
value of  $\hat{m}_s$, the average of these different numbers and its
relative error the one coming from the most accurate determination
around 1.2 GeV. The reason for this choice of the error is mainly due to
the fact that the average would have been
obtained in the $compromise$ region around 1.2--1.3 GeV, where
the higher state mesons
and the non-perturbative contributions, which are the main
sources of uncertainties are both small, while for lower (higher)
energies the uncertainties due to the non-perturbative (higher state
mesons) are large.
\begin{table*}[t]
\setlength{\tabcolsep}{1.5pc}
\caption{Estimates of  $\hat{m}_s$ and of $\bm_s$(1 GeV) to
three-loops
from $\Delta_{10}$}
\begin{tabular}[t]{c|c c c c  }
\hline
&&&& \\
&\multicolumn{4}{c}{ $\hat{m}_S$ to three-loop accuracy}  \\
&&&& \\
$\Lambda_3$[MeV]&290&375&480&540\\
&&&& \\
\hline
&&&& \\
$M_{\tau}$[GeV]&&&& \\
$1.0$  &$208.7\pm 43.8$&$155.8 \pm 32.7$&$94.00 \pm 19.7$&$59.10
\pm 12.4$\\
$1.2$ &$222.2\pm 28.0$&$179.5 \pm 22.6$&$131.3 \pm 16.5$&$104.4
\pm 13.2$\\
$1.4$ &$177.1\pm 35.8$&$149.2 \pm 30.2$&$118.1 \pm 23.9$&$101.2
\pm 20.4$\\
$1.6$ &$169.9\pm 47.6$&$146.5 \pm 41.0$&$121.1 \pm 33.9$&$107.5
\pm 30.1$\\
&&&& \\
 $\la\hat{m}_S\ra $&$200.6\pm 25.4$& $163.2 \pm 20.5$
&$116.9\pm 14.7$&
$85.4\pm 10.8$\\
&&&& \\
\hline
&&&& \\
$\bm_s(1~\mbox{GeV})$&$208.7\pm 26.3$&$196.7\pm 24.7$
&$176.9\pm 22.3$&$156.7\pm 19.7$\\
&&&& \\
\hline
\end{tabular}
\end{table*}
\nin
Using the previous value of $\hat{m}_s$ and Eq. (20),
 we finally deduce in Table 3 the
value of the running mass evaluated at 1 GeV to three-loops.
One should notice that if we have used the usual $a_s$-expansion
but not the one of \cite{LDP} for
$R_{\tau,I}$, we would have obtained  value of about
7$\%$ higher. We consider this difference as intrinsic systematic
uncertainties of the approach.
\nin
Therefore, for $\alpha_s(M_Z)=0.118\pm 0.006$ \cite{BETHKE,PDG}
which corresponds to
order $a^2_s$ to $\Lambda_3=375^{+105}_{-85}$ MeV, we deduce
including the $a^2_s$-corrections:
\beq
\overline{m}_s(1~\mbox{GeV})= (196.7 \pm 24.7 \pm 14.8 \pm 4.0)~
\mbox{MeV},
\eeq
where the first error is due to the data and to the dimension-six
condensates which contribute about 50$\%$ each, the second one is due
to the choice of the order parameter in the expansion of perturbative
series and to a guess of the unknown higher order terms, while the last
small error is due to the value of $\Lambda_3$. Therefore,
we deduce as a $final$ estimate:
\beq
\overline{m}_s(1~\mbox{GeV})= (196.7 \pm 29.1)~
\mbox{MeV},
\eeq
\nin
One can also notice that the inclusion of
the $a^2_s$-terms has reduced by about 10$\%$ the two-loop result.
\nin
We consider this result as an improvement of the earlier
inaccurate results
\cite{REIND,SNB} from the $\phi$-meson sum rule. Moreover,
as pointed out in \cite{SNB}, the sum rule analysis
suffers from the competition between the
quark mass corrections and the dimension four-quark condensate
contribution linear in $m_s$ where the uncertainties
in the value of $\la \bar ss \ra$
and gluon $\la \alpha_s G^2\ra$ condensates mask the major part of
the quark mass corrections.
\nin
This value is
in agreement within the errors with the vailable estimates
from  the (pseudo) scalar sum rules \cite{SNB},
\cite{PSEUDO}-\cite{JAMIN}
\footnote{However these results depend on the appreciations of the high
meson mass contributions to the spectral function (see e.g. \cite{KNECHT}
for a different realization of chiral symmetry not used in the previous
papers).} and from lattice calculations \cite{LATT}
\footnote{If one uses the most recent value from the scalar sum rule
\cite{JAMIN}:
$\overline{m}_s(1~\mbox{GeV})=(189 \pm 32)~ \mbox{MeV}$,
and the lattice result
\cite{LATT} rescaled at 1 GeV: $
\overline{m}_s(1~\mbox{GeV})=(180\pm 25\pm 25) \mbox{MeV}$,
where the last error is a conservative guess
of the systematic uncertainties
of the approach, it is informative to deduce the $average$:
$\overline{m}_s(1~\mbox{GeV})= (190.3 \pm 18.4)~ \mbox{MeV}$.}.
\section{Implications for the $\chi SB$ parameters}
\subsection*{Values of $\overline{m}_d$, $\overline{m}_u$
and of the associated condensates}

\nin
In the following, we combine the previous value of
$\overline{m}_s$ from the vector channel with the value of the
sum of the up and down quark masses determined recently from the
pseudoscalar sum rule to three-loop accuracy \cite{BIJ}:
\beq
(\overline{m}_d+\overline{m}_u)\ga{1~\mbox{GeV}}\dr=
(12.0 \pm 2.5)~\mbox{MeV}.
\eeq
One can deduce:
\beq
r_3\equiv \frac{m_s}{1/2(m_u+m_d)}=32.8 \pm 8.4,
\eeq
which is compatible within the errors, but slightly
higher than the standard current algebra determination \cite{GASSER}:
\beq
r^{CA}_3\equiv \frac{m_s}{1/2(m_u+m_d)}=25.7 \pm 2.6.
\eeq
Although less accurate, our result
is not affected by the uncertainties
of the current algebra value
pointed out in \cite{MANO}, but, as mentioned
previously, the result from the pseudoscalar
sum rule relies strongly on the parametrization of the 3$\pi$
contribution to the spectral function which depends on the realization
of chiral symmetry due to the absence of the data for this channel.
\nin
We combine the previous value of $\overline{m}_s$
with the current algebra relation \cite{GASSER}
\footnote{One could instead use the value of the quark mass difference
determined from the scalar sum rule \cite{SCAL,PSEU}, but this channel
is not well-known  experimentally as the true nature of the
$a_0(980)$ is not yet well-understood.}:
\beq
r_2^{CA}\equiv\frac{(m_d-m_u)}{(m_d+m_u)}=
\frac{m_\pi^2}{M_K^2}\frac{(M^2_{K^0}-M^2_{K^+})_{QCD}}{M_K^2-m^2_\pi}
\frac{m^2_s-\hat{m}^2}{(m_u+m_d)^2}=(0.52\pm 0.05)10^{-3}(r_3^2-1),
\eeq
where $2\hat{m}=m_u+m_d$; the QCD part of the $K^+-K^0$ mass-difference
comes from the estimate
of the electromagnetic term using the
Dashen theorem including next-to-leading chiral corrections \cite{BIJ2}.
Using our previous value of $r_3$ in Eq. (36) and
the pseudoscalar sum rule result, we obtain to three-loop accuracy:
\beq
(\overline{m}_d-\overline{m}_u)\ga{1~\mbox{GeV}}\dr=
(6.7 \pm 3.4)~\mbox{MeV}.
\eeq
Using again the pseudoscalar sum rule result, we finally obtain:
\beq
\overline{m}_d(1~\mbox{GeV})=(9.4\pm 2.1)~\mbox{MeV}, ~~~~~~~
\overline{m}_u(1~\mbox{GeV})=(2.7\pm 2.1)~\mbox{MeV}.
\eeq
If we use, instead, the previous values of the ratios $r^{CA}_3$
and $r^{CA}_2$ from
current algebra and the previous value of $m_s$, we obtain:
\beq
(\overline{m}_d+\overline{m}_u)\ga{1~\mbox{GeV}}\dr=
(15.3 \pm 2.7)~\mbox{MeV},~~~~~
(\overline{m}_d-\overline{m}_u)\ga{1~\mbox{GeV}}\dr=
(5.2 \pm 1.4)~\mbox{MeV},
\eeq
and then:
\beq
\overline{m}_d(1~\mbox{GeV})=(10.3\pm 1.5)~\mbox{MeV}, ~~~~~~~
\overline{m}_u(1~\mbox{GeV})=(5.0\pm 1.5)~\mbox{MeV},
\eeq
showing that the two alternative approaches lead to consistent
 values of $m_d$ and $m_u$, though one
can notice that the value of $m_d$ is almost unaffected by the
change of the ratio
$r_3$ once $r_2$ is given, while $m_u$ becomes smaller for a larger
value of $r_3$.
These values are in good agreement with the previous
estimates \cite{PSEUDO,SNB,SNP,BIJ}, which have been, however,
obtained for smaller values of
$\Lambda$. The reason is that, contrary to the
invariant mass $\hat{m}_s$, the running mass is not very sensitive to
the $\Lambda$-values in the (pseudo)scalar sum rules analysis.
We consider as a final estimate the average of the two previous
results:
\beq
\overline{m}_d(1~\mbox{GeV})=(10.0\pm 1.2)~\mbox{MeV}, ~~~~~~~
\overline{m}_u(1~\mbox{GeV})=(4.2\pm 1.2)~\mbox{MeV}.
\eeq
Using the pion PCAC relation in Eq. (24), one can deduce:
\beq
\frac{1}{2}~\la \bar uu + \bar dd \ra~(1~\mbox{GeV})
= -[(228.6\pm 9.1)~\mbox{MeV}]^3.
\eeq
\subsection*{$K'(1.46)$ and $K^*_0(1.43)$ decay constants}
 An earlier QCD spectral sum rule analysis of the
(pseudo)scalar current \cite{SNK,SNP,SNB} has provided an estimate of
the decay constants of the $K'(1.46)$ and $K^*_0(1.43)$ mesons in the
scheme where the spectral function is represented by the usual duality
ansatz: lowest narrow resonances plus a QCD continuum from a threshold
$t_c$.
In our approach the decay constant will be an {\it effective coupling}
including into it $all$ corrections due to finite widths and to
threshold effects. For our purpose, we come back again to the familiar
(pseudo)scalar sum rule \cite{SNB}-\cite{BIJ} in
the strange quark channel, where we shall consider
the Laplace sum rule:
\beq
{\cal{L}}_0\equiv \int_{0}^{\infty}dt~e^{-t\tau}~
\frac{1}{\pi}~\mbox{Im}\Psi_{(5)}(t)^u_s=
\frac{N_c}{8\pi^2}(\overline{m}_s\pm\overline{m}_u)^2 \tau^2\aga
1+\sum_{D=0,2,...}\delta^{(D)}_0{(5)}\adr,
\eeq
where $\Psi_{(5)}(q^2)^u_s$ is the two-point correlator associated to
the (pseudo)scalar current:
\beq
\partial_\mu A(V)^\mu (x)^u_s=(m_u\pm m_s) :\bar u (i\gamma_5)s:.
\eeq
$\delta^{(D)}_0{(5)}$ are the (non)perturbative corrections of the
correlator. They read:
\bea
\delta^{(0)}_0{(5)}&=& 4.82a_s -9.69a_s^2 \nnb \\
\delta^{(2)}_0{(5)}&=& -2\tau\overline{m}_s^2(1+6.87a_s)\nnb \\
\delta^{(4)}_0{(5)}&=&\tau^2\frac{\pi}{3}\aga \la\alpha_s G^2\ra -4\pi
 m_s(\la \bar uu \mp\bar ss \ra)\adr \nnb \\
\delta^{(6)}_0{(5)}&=&\frac{896}{81}\pi^3\rho \alpha_s \la \bar uu
\ra ^2.
\eea
 We use the duality ansatz parametrization of the pseudoscalar spectral
function with the two resonances $K,
{}~K'(1.46)$ in the sum rule ${\cal{L}}_0$ for
extracting the $K'(1.46)$ decay constant which is controlled by the
ratio:
\beq
r_K\equiv {M^4_{K'}f_{K'}^2}/{M_K^4f_K^2}.
\eeq
Using $m_s$ and the correlated values of $\Lambda_3$ given in Table 3,
we study, as usual, the stability of the result with respect to the sum
rule variable $\tau$ and the QCD continuum threshold $t_c$. The
$\tau$-dependence of the sum rule is quite flat in a large range of
$\tau$ smaller than 0.8 GeV$^{-2}$
\footnote{The flatness of the sum rule prediction in this large
range of $\tau$-values can also be interpreted as a strong
evidence for the negligible effect of the instanton
effects for $M\equiv 1/\sqrt{\tau} \geq 1.2$ GeV.},
 while the optimal result corresponds to
$\tau \simeq 0.2$ GeV$^{-2}$ and for the QCD continuum threshold $t_c$
of about $6\sim 7$ GeV$^2$, a set of parameters consistent with the one
obtained recently \cite{JAMIN}.
Taking into account the correlations
between $m_s,~\Lambda_3,~t_c$ and $r_K$, we deduce:
\beq
r_K = 7.0\pm 2.5,
\eeq
which we consider as an update of the result $7\pm 1$ obtained in \cite
{SNB,SNP}.
\nin
We perform a similar analysis in the scalar channel in order to fix
the decay constant of the $K^*_0(1.43)$ resonance and the correlated
value
of the QCD continuum. The optimal stability is again
obtained for $t_c\simeq 6\sim 7$ GeV$^2$ and for $\tau \simeq 0.4$
GeV$^{-2}$. We obtain:
\beq
f_{K^*_0}\simeq (40.2\pm 6.2)~\mbox{MeV},
\eeq
which is again an update of the result in \cite{SNB,SNP}.
\subsection*{Test of kaon PCAC and value of $\la \bar ss \ra /
\la \bar uu \ra$}
\nin
We use the previous results in order to extract the value of the
subtraction constants $\Psi_{(5)}(0)^u_s$ from which we can
test the deviation from kaon PCAC. Using as in \cite{SNP,SNB},
the Laplace sum rule ${\cal{L}}_{-1}$
obeyed by these quantities:
\beq
{\cal{L}}_{-1}\equiv \int_{0}^{\infty}\frac{dt}{t}~e^{-t\tau}~
\frac{1}{\pi}~\mbox{Im}\Psi_{(5)}(t)^u_s=\Psi_5(0)^u_s+
\frac{N_c}{8\pi^2}(\overline{m}_s\pm\overline{m}_u)^2 \tau\aga
1+\sum_{D=0,2,...}\delta^{(D)}_{-1}{(5)}\adr,
\eeq
where $\delta^{(D)}_{-1}{(5)}$ are the (non)perturbative corrections,
we find that the analysis
is not conclusive as the result does not have a clear stability and
increases with $\tau$.
We therefore use the FESR proposed in \cite{PSEU}. Expressing:
\beq
\Psi_5(0)^u_s=2M_K^2f_K^2(1-\delta_K),
\eeq
one has the sum rule \cite{PSEU}:
\beq
\delta_K\simeq \frac{3}{16\pi^2}\frac{\overline{m}_s^2t_c}{f^2_KM^2_K}
\aga 1+\frac{23}{3}a_s + {\cal{O}}(a^2_s)\adr
-r_K\ga\frac{M_K}{M_{K'}}\dr^2,
\eeq
which gives, after using the {\it correlated values} of the input
parameters:
\beq
\delta_K= 0.34^{+0.23}_{-0.17},
\eeq
confirming a large violation of kaon PCAC \cite{SNP,SNB}. In terms of
the
{\it normal-ordered} condensates, one has:
\beq
\Psi_{(5)}(0)^u_s=-(m_u\pm m_s)\la \bar ss \pm \bar uu \ra.
\eeq
Therefore, we deduce
\footnote{We have checked that an extraction of this ratio from the
$\phi$
meson sum rule is not conclusive.}:
\beq
{\la \bar ss \ra}/{\la \bar uu \ra} = 0.71^{+0.59}_{-0.42}.
\eeq
One can do a similar analysis for the scalar channel. The corresponding
FESR is:
\beq
\Psi(0)^u_s=2M_{K^*_0}^2f_{K^*_0}^2-
\frac{3}{16\pi^2}{\overline{m}_s^2t_c}
\aga 1+\frac{23}{3}a_s + {\cal{O}}(a^2_s)\adr,
\eeq
which gives:
\beq
\Psi(0)^u_s=-\ga 7.8^{+5.5}_{-2.7}\dr 10^{-4}~\mbox{GeV}^4,
\eeq
in agreement with previous results \cite{SNB,SNP}. Taking the ratio of
the
scalar over the pseudoscalar subtraction constants, one can deduce:
\beq
{\la \bar ss \ra}/{\la \bar uu \ra} = 0.68^{+0.15}_{-0.29},
\eeq
which we consider as an update of the previous results in \cite{SNP,SNB}.
This result supports the $SU(3)$ breaking of the
condensates from the baryon sum rules \cite{DOSCH} and
agrees with the one from chiral
perturbation theory \cite{GASSER} around 0.72$\sim$0.76.
If one instead works with the {\it non-normal ordered} condensate, one
should add to the expression in Eq. (55) a small perturbative piece
first obtained by Becchi et al \cite{PSEUDO} (see also
\cite{SNB,BROAD,JAMIN}):
\beq
\Delta_P=-\frac{3}{2\pi^2}\frac{2}{7}\ga \frac{1}{a_s}-\frac{53}{24}\dr
\overline{m}_s^4.
\eeq
This leads to the ratio of the {\it non-normal ordered} condensates:
\beq
{\la \bar ss \ra}/{\la \bar uu \ra} = 0.87^{+0.03}_{-0.28}.
\eeq
\subsection*{Determination of the ratio $m_b/m_s$}
\nin
Finally, we combine the previous value of $m_s$ with the running mass of
the $b$-quark evaluated to two-loop accuracy from the $\Upsilon$
sum rules \cite{SNM}.
\nin
For consistency, we use the two-loop version of the result in
Eq. (34), which corresponds to the two-loop running mass:
\beq
\overline{m}_s(1~\mbox{GeV})= (222 \pm 22)~ \mbox{MeV}.
\eeq
In so doing, we run the strange quark mass until the $b$-pole mass
where the $b$-quark one has been evaluated. We take care on the different
threshold effects by using the relation between the running mass
evaluated for a flavour $f$ and $f$-1 \cite{BERN}:
\beq
\overline{m}_i^{(f-1)}(\nu)=\overline{m}_i^{(f)}(\nu)
\aga 1+\frac{1}{12}\ga x^2+\frac{5}{3}x+\frac{89}{36}\dr a_s^2(\nu)\adr,
\eeq
where $x\equiv 2\ln{M(\nu)/\nu}$; $M(\nu)$ is the mass of the excited
heavy quark at the matching point. At the two-loop accuracy where the
$b$-quark running mass has been estimated \cite{SNM}, one can have:
\beq
\overline{m}_i^{(f-1)}(\nu)=\overline{m}_i^{(f)}(\nu)\aga 1 +
{\cal{O}}(\alpha_s^2)\adr
\eeq
at the heavy quark thresholds. We use this relation together with the
matching conditions for the coupling constant \cite{BERN,SANTA}:
\beq
\alpha^{(f-1)}_s=\alpha^{(f)}_s+ {\cal{O}}(\alpha_s^3).
\eeq
We take to two-loop accuracy \cite{SNM}:
\beq
\overline{m}_c^{(4)}(M_c)= (1.23\pm 0.05)~\mbox{GeV},~~~~~~~~~~~~
\overline{m}_b^{(5)}(M_b)= (4.23\pm 0.04)~\mbox{GeV},
\eeq
where the index (4), (5) indicates the number of excited flavours
and we use \cite{SNM}:
\beq
M_c=(1.42\pm 0.03)~\mbox{GeV},~~~~~~~~~~~~
M_b=(4.62\pm 0.02)~\mbox{GeV}.
\eeq
We multiply the quoted errors by a factor 10 in order to have a large
overestimate of the error in the following analysis.
Then, we deduce the running strange quark mass for 5 flavours at $M_b$:
\beq
\overline{m}_s^{(5)}(M_b)= (125\pm 15)~\mbox{MeV},
\eeq
where the errors due to the thresholds are much smaller than the ones
induced by the determination of $m_s$.
By combining this result with
$\overline{m}_b^{(5)}(M_b)$ in Eq. (66),
one can deduce the scale independent mass ratio:
\beq
r_5\equiv {m_b}/{m_s}=34\pm 4,
\eeq
which is an useful quantity for model-buildings, and where
it is amusing to notice that we have $r_3\approx r_5$!
\section{Conclusion}
We have estimated the running strange quark mass of the
$\overline{MS}$-scheme using a tau-like decay-modern version of the old
DMO sum rule relating the $I=0$ and $I=1$ component of the $e^+e^-\rar$
hadrons total cross-section, which, contrary to the existing approaches,
has the $great$ advantage to be {\it model-independent} and to be free
from the less-controlled instanton  contributions.
 Our $final$ result is given in Eq. (34).
\nin
By combining this
result with the existing estimate from the pseudoscalar
sum rules and/or current algebra
based on the $standard$ realization of chiral symmetry (dominance
of the linear term in the quark mass expansion of the pseudoscalar meson
mass squared),
we have deduced the value of the up and down running masses in Eq. (43)
and the values of the quark condensates in Eqs. (44).
\nin
We use our previous value of $m_s$ into the (pseudo)scalar sum rules in
order to extract without theoretical prejudices the decay constants of
the $K'(1.46)$ and $K^*_0(1.43)$ mesons which can absorb into them all
possible
different hadronic corrections to the spectral functions. Our results
are given in Eqs. (49) and (50) and confirm previous findings from
QCD spectral sum rules \cite{SNP,SNB}.
\nin
We use the previous results in order to extract the size of the
(pseudo)scalar two-point function subtraction constants. As a
consequence,
we obtain a deviation of about 40\% from kaon PCAC (Eq.(54)) and a
large $SU(3)$ breaking of the ratio of the {\it normal-ordered} quark
condensates (Eq. (59)), which confirm again the QSSR results in
\cite{SNP,SNB}.
\nin
Finally by combining the strange quark result with the estimate of
the running $b$-quark
mass determined directly from the $\Upsilon$ sum rules, we deduce
the ratio $r_5\equiv m_b/m_s=34\approx r_3\equiv 2m_s/(m_u+m_d)$ given
in Eq. (69). We expect that the almost equal value between $r_5$ and
$r_3$ is not only an accident of nature but may be due to a richer
symmetry which could be explained from the theory of
unification of interaction forces.
\section*{Acknowledgements}
It is a pleasure to thank A. Pich for exchanges and for carefully
reading the manuscript.
\vfill \eject

\end{document}